\documentclass[prl,amsmath,superscriptaddress,amssymb,twocolumn,floatfix,aps,showpacs,showkeys]{revtex4-1}
\usepackage{amssymb}   
\usepackage{amssymb}
\usepackage{epsfig}
\usepackage{graphicx} 
\usepackage[space]{grffile}
\usepackage{color}

\def \figwidth{8.4cm}

\begin{document}
\title{Breit-Wigner phase is a fundamental property of a resonance}

\author{S.~Ceci}
\email{sasa.ceci@irb.hr}
\affiliation{Rudjer Bo\v{s}kovi\'{c} Institute, Bijeni\v{c}ka  54, HR-10000 Zagreb, Croatia}
\author{M.~Vuk\v si\' c}
\affiliation{University of Zagreb, Bijeni\v{c}ka  34, HR-10000 Zagreb, Croatia}
\author{B.~Zauner}
\affiliation{Rudjer Bo\v{s}kovi\'{c} Institute, Bijeni\v{c}ka  54, HR-10000 Zagreb, Croatia}


\begin{abstract}
In the course of devising a simple method for extraction of the S-matrix poles from the data, an additional fundamental resonance property emerged. It is a reaction invariant quantity, and since it is directly related to the Breit-Wigner parameters, we call it the Breit-Wigner phase $\beta$. We propose that this $\beta$ is added in resonant data tables. 
\end{abstract}


\maketitle

\section{Introduction}

In this year's {\em Review of particle physics} \cite{PDG}, resonances are defined by poles of the S-matrix, whether in scattering, production, or decay matrix elements. Positions of the S-matrix poles are, as stated there, independent of the process. Therefore, the poles are the fundamental physical properties of the S-matrix. As such, they are in contrast to other quantities related to resonance phenomena like Breit-Wigner parameters or K-matrix poles.

In the same reference two strong statements are made that succinctly summarize the current understanding of the physical properties of resonances. The first is that the Breit-Wigner parameters depend on the formalism used (angular-momentum barrier factors, cut-off parameters, and the assumed or modeled background). The other is that the accurate determination of pole parameters from the analysis of data on the real energy axis requires the implementation of the correct analytic structure of the relevant (often coupled) channels.

In this Letter we show that neither of the statements is entirely correct. We propose a simple few-parameter model to extract pole parameters directly from the data, without details on the analytic structure of the S-matrix, and extract the Breit-Wigner parameters independently of any formalism and knowledge of the barrier factors, cut-off parameters or background. In addition, we show that the Breit-Wigner parameters are, along with the pole parameters, the key properties of the S-matrix. 

This result is especially important because {\em ab initio} hadronic models, such as the lattice quantum chromodynamics \cite{Dur08} and various quark models (see e.g., Ref.~\cite{QM}), often use Breit-Wigner parameters. If those parameters were formalism dependent, the predictions would be unreliable. 

Some results in the literature strongly suggest that the Breit-Wigner parameters are fundamentally flawed. In Ref.~\cite{Scherer} the authors show that the Breit-Wigner parameters of $\Delta(1232)$ change under a particular field transformation which leaves other physical properties, including the poles of the S-matrix, invariant. Moreover, in Ref.~\cite{Sirlin} a similar feature is shown for the Breit-Wigner parameters of the Z boson. 

Here we show that the definition of the Breit-Wigner parameters used in Refs.~\cite{Scherer,Sirlin} was not appropriate and that the conclusion on non-physicality of those parameters is not valid. Unfortunately, that was not the last of the problems with Breit-Wigner parameters. 

There is a lot of confusion in the literature regarding the Breit-Wigner mass. Sometimes it is mistaken by the bare mass as in Ref.~\cite{Liu06} (see Ref.~\cite{Cec09}). Sometimes it is used in models where the pole mass should be used instead as in Ref.~\cite{DeCruz12} (cf.~the original model in Ref.~\cite{DC12model}). And sometimes resonant peak positions are confused for the pole positions as was done in Ref.~\cite{BES06}. That last paper led us to look into the connection of the peaks with pole and Breit-Wigner parameters. In Ref.~\cite{Cec13} we have shown that the resonant peak positions are often much closer to the Breit-Wigner than to the pole masses but, generally, they are neither. 

Here we explain the interrelation between these three resonant features and show that there is a surprisingly simple relation between its Breit-Wigner mass, pole position, and the peak position.

However, the biggest surprise with the proposed model is related to the K-matrix. In Ref.~\cite{CecPLB08} we have analyzed a well-known unitary analytic coupled-channel approach \cite{CMB} and noted that the extracted K-matrix poles are almost always similar to the Breit-Wigner masses. Moreover, each pole was at the same energy as the peak of the imaginary part and the zero of the real part of the amplitude-matrix trace. Interestingly enough, when we calculate the K-matrix using the unitarized version of the first-order non-analytic model introduced in this Letter, we obtain exactly the same features as we saw in Ref.~\cite{CecPLB08}. The Breit-Wigner mass is indeed the K-matrix pole. Since the K-matrix is uniquely defined by the S-matrix, the Breit-Wigner mass is a fundamental property of the S-matrix. 

\section{Model}

In case that particle spins are not measured, the formula for the resonant cross section is given by
\begin{equation}\label{BreitWignerFormula}
\sigma = \frac{4\pi}{q^2} \, \frac{2J+1}{(2s_1+1)(2s_2+1)} \,\left|A\right|^2+ \sigma_\mathrm{bg},
\end{equation} 
where $q$ is c.m.~momentum of incident particles, $J$ is the spin of the resonance, $s_1$ and $s_2$ are the spins of incident particles, and $\sigma_\mathrm{bg}$ is a contribution of background processes. The key object in this relation is a resonant amplitude $A$, generally written as
\begin{equation}\label{MoreRealisticAmplitude}
A = \frac{V\mathrm(W)}{m_0-W-i\,\Sigma(W)},
\end{equation} 
where $W$ is c.m.~energy, the vertex function $V$ is a real, and the self energy $\Sigma$ is a complex function of energy $W$.

We get the familiar {\bf Breit-Wigner formula} \cite{BW} by assuming that both $V$ and $\Sigma$ are real-valued and do not change much with energy
 \begin{equation}\label{OriginalBWAmplitude}
A\approx \frac{\Gamma_\mathrm{par}/2}{m_0-W-i\,\Gamma_\mathrm{tot}/2}.
\end{equation} 
Here we use the standard terminology in this field: $\Gamma_\mathrm{par}$ is the partial decay width, i.e., the inverse of the resonance mean lifetime that gives the rate at which a resonance decays into the observed channel. In natural units, it has the same unit as the mass $m_0$. $\Gamma_\mathrm{tot}$ is the total decay width, i.e., the sum of all partial decay widths. This form is not suitable for broad resonances nor resonances with substantially deformed shape. It should be pointed out that, in spite of this formula's name, the parameters $m_0$ and $\Gamma_\mathrm{tot}$ are generally not the Breit-Wigner parameters.  
 
The {\bf Breit-Wigner parameters} are usually obtained from a more realistic Flatt\'e formula \cite{Flatte}, where both $V$ and $\Sigma$ are assumed to be real functions with particular properties. These properties may vary and this is the reason why the extracted Breit-Wigner parameters often differ from one analysis to another, even if the same dataset is analyzed. Such a formula was used in Ref.~\cite{Liu06} and produced a wrong result because in the analyzed energy region $\Sigma$ was complex. 

Contrary to the aforementioned approaches, in our model we acknowledge the fact that $\Sigma$ is a complex and $V$ is a real function. Expanding them to the first order, we get a general form 
\begin{equation}\label{LinearAmplitude}
A = \frac{|r| \, e^{i\theta} }{M-W-i\, \Gamma/2}+A_\mathrm{B} \, e^{i\beta},
\end{equation}
where again we use the standard terminology, but this time for the {\bf resonant pole parameters}: mass $M$, total decay width $\Gamma$, residue magnitude $|r|$, and residue phase $\theta$. In addition, there are two real parameters: the background $A_\mathrm{B}$ and its phase $\beta$. 

Since $V$ is a real function, these parameters are not independent, which allows us to eliminate $A_\mathrm{B}$  
\begin{equation}
A_\mathrm{B} = \frac{2\, |r|}{\Gamma}\, \sin(\theta-\beta).
\end{equation} To simplify the notation, we introduce the branching fraction $x = 2\, |r| \,  / \, \Gamma$ and the phase difference 
\begin{equation}
\delta = \theta-\beta \label{deltaDefinition}.
\end{equation}
Putting it all together,  the amplitude in Eq.~(\ref{LinearAmplitude}) becomes
\begin{equation}\label{AmplitudeFinal}
\boxed{A=  x \, e^{i\left(\rho+\beta\right)} \, \sin\left(\rho+\delta\right),}
\end{equation}
where the resonant phase shift $\rho$ is defined as
\begin{equation}\label{rhoDefinition}
\tan \rho = \frac{\Gamma/2}{M-W}.
\end{equation}
Basically, this is an {\bf improved Breit-Wigner formula} with five independent parameters: $M$, $\Gamma$, $|r|$, $\theta$,  and the new parameter $\beta$, which we call the Breit-Wigner phase. This new phase parameter $\beta$ is the difference between the amplitude's phase at the energy equal to the pole mass $M$ and $90^\circ$.

\section{Mass formulas}

In Ref.~\cite{Cec13} we have phenomenologically obtained a relation for $|A|^2$, which is in full agreement with the result we get here by squaring the magnitude of Eq.~(\ref{AmplitudeFinal}). The peak position of $|A|^2$, labeled here as $M_\mathrm{mag}$, is given by 
\begin{equation}\label{Mpeak}
M_\mathrm{mag} = M - \frac{\Gamma}{2}\,\tan\delta. 
\end{equation}
It is, therefore, no surprise that this is exactly the same formula we have got in Ref.~\cite{Cec13}. Next, we propose a few additional mass formulas derivable from Eq.~(\ref{AmplitudeFinal}). 

We obtain the second mass formula by analyzing the imaginary part of $A$ and noting that it has a peak at some energy, which we call $M_\mathrm{Im}$, with the value
\begin{equation}\label{MIm}
M_\mathrm{Im} = M - \frac{\Gamma}{2}\,\tan\frac{\theta}{2}.
\end{equation}
This formula could be very useful because due to the optical theorem, the imaginary part of the elastic amplitude is related to the total cross section $\sigma_\mathrm{tot}$. In fact, we get $\sigma_\mathrm{tot}$ by replacing $|A|^2$ with $\mathrm{Im}\, A$ in Eq.~(\ref{BreitWignerFormula}).

The third and arguably the most important mass formula is obtained by linear expansion of the numerator and denominator in Eq.~(\ref{MoreRealisticAmplitude}), using the standard definition of the {\bf Breit-Wigner mass} as the zero of the real part of the denominator in Eq.~(\ref{MoreRealisticAmplitude}) (e.g., see Refs. \cite{Scherer,Sirlin,Man95})
\begin{equation}
\boxed{M_\mathrm{BW} = M - \frac{\Gamma}{2}\, \tan\beta.\label{BWMass}}\\
\end{equation}

The {\bf Breit-Wigner width} is then given by \mbox{$\Gamma_\mathrm{BW} = \Gamma/\cos^2\beta$}, which is a useful relation for consistency checking. Furthermore, note that $M_\mathrm{BW}$ is also the peak of $(\mathrm{Im}\, A)^2/|A|^2$ and that it is given by the energy at which \mbox{$\rho+\beta$}, the phase of $A$, crosses $90^\circ$. 


Before proceeding any further we need to address the issue of the {\bf Breit-Wigner mass non-physicality} pointed out by Scherer et al.~in \cite{Scherer} and Sirlin in \cite{Sirlin}. They have shown that a quantum field transformation, which does not change observables nor pole parameters, will change the Breit-Wigner mass if it is defined as the zero of the real part of the denominator in Eq.~(\ref{MoreRealisticAmplitude}). However, since the transformation changes the denominator but does not change the amplitude itself (otherwise it would change the observables as well), we can consistently redefine the Breit-Wigner mass using the whole invariant resonant amplitude, both the denominator and numerator, and it will be invariant to the transformation: {\bf the Breit-Wigner mass is the energy at which the real part of the resonant amplitude becomes zero}, i.e., when its phase crosses $90^\circ$. Thus, Eq.~(\ref{BWMass}) is still valid. 

To test the proposed mass formulas, we use them on a familiar example: the Z boson. The resonant amplitude of the Z boson is given by \cite{PDG} 
\begin{equation}\label{ARZboson}
A_\mathrm{Z} = \frac{x\, \Gamma_\mathrm{BW}\,W^2/M_\mathrm{BW}}{M_\mathrm{BW}^2-W^2-i\,\Gamma_\mathrm{BW}\,W^2/M_\mathrm{BW}}.
\end{equation}
All parameters in this relation are the Breit-Wigner parameters. To test the formulas, we extract the pole parameters $M$, $\Gamma$, and $\theta$ directly from this relation. Then we use the mass formulas to reproduce the Breit-Wigner parameters. 

First we note that, by construction, the peak of $\mathrm{Im}\,A$ is at the same position as the peak of $|A|^2$. Therefore $\delta$ must be equal to $\theta/2$, as is evident by comparison of Eqs.~(\ref{Mpeak}) and (\ref{MIm}). In addition, since $\beta$ is $\theta-\delta$, it will also be equal to $\theta/2$. Finally, we calculate the Breit-Wigner parameters using Eq.~(\ref{BWMass}). The agreement between the original Breit-Wigner parameters from Ref.~\cite{PDG} and those obtained by Eqs.~(\ref{Mpeak}-\ref{BWMass}) turn out to be excellent, as can be seen in Table \ref{ZbosonTable}. 

\begin{table}[h!]
\caption{Agreement between the PDG values of the Z-boson resonant parameters and those obtained by using the mass formulas (\ref{Mpeak}-\ref{BWMass}) applied to the Z-boson amplitude (\ref{ARZboson})  \label{ZbosonTable}}
\begin{ruledtabular}
\begin{tabular}{lllcc}
{\bf Z boson}	& Mass & Width & $\theta$ & $\beta$\\
Source 		& (MeV) & (MeV) & ($^\circ$) & ($^\circ$)\\
\hline
BW PDG \cite{PDG}	&	 	91 187.6$\pm$2.1  & 2495.2$\pm$2.3 & N/A &\\
Pole PDG \cite{PDG} &		91 162 &    N/A &  & N/A\\
\hline
Pole Eq.~(\ref{ARZboson}) &	91 162.0 & 2494.0  & $-2.4$   &   \\
BW Eq.~(\ref{BWMass}) & 91 187.6 & 2495.1 &   & $-1.2$		\\
\end{tabular}
\end{ruledtabular}
\end{table}

\section{Resonant parameters from data}

In practice, the resonant parameters are extracted from the scattering data. Therefore, we need to test how well the proposed model fits the data and how well the extracted parameters agree with the known values. We choose $\Delta(1232)$ because it has, like the Z boson,  large and statistically significant difference between the Breit-Wigner and the pole mass (cf. Table \ref{Delta1232Table}). 

\begin{figure}[h!]
\includegraphics[width=\figwidth]{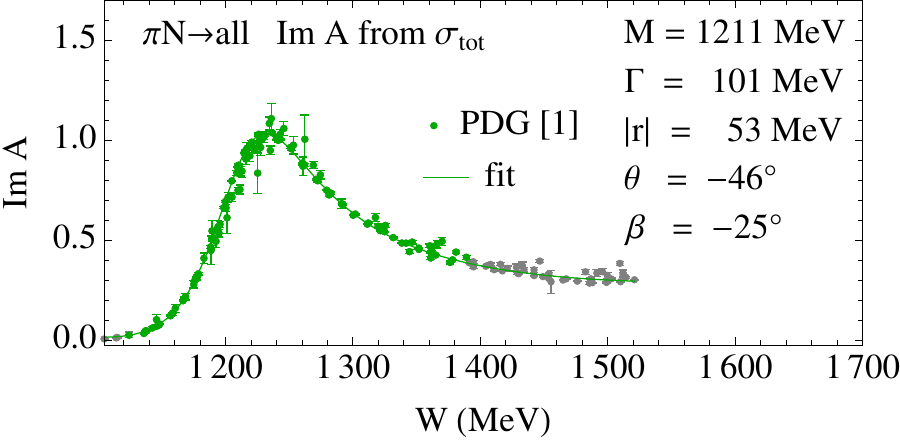}
\caption{ Fit of Eq.~(\ref{AmplitudeFinal}) to the $\mathrm{Im}\,A$ data obtained from the total cross section $\sigma_\mathrm{tot}$ for $\pi N$ scattering using Eq.~(\ref{BreitWignerFormula}). The data is taken from the PDG database \cite{PDG}. The fitting strategy and background treatment are the same as in Ref.~\cite{Cec13}  \label{fig2}}
\end{figure}

First, in Fig.~(\ref{fig2}), we show the fit of Eq.~(\ref{AmplitudeFinal}) to the $\mathrm{Im}\,A$ data obtained from the total cross section $\sigma_\mathrm{tot}$ for $\pi N$ scattering using Eq.~(\ref{BreitWignerFormula}). The data is taken from the PDG database \cite{PDG}, while the fitting strategy and background treatment are the same as in Ref.~\cite{Cec13}. 

\begin{figure}
\includegraphics[width=\figwidth]{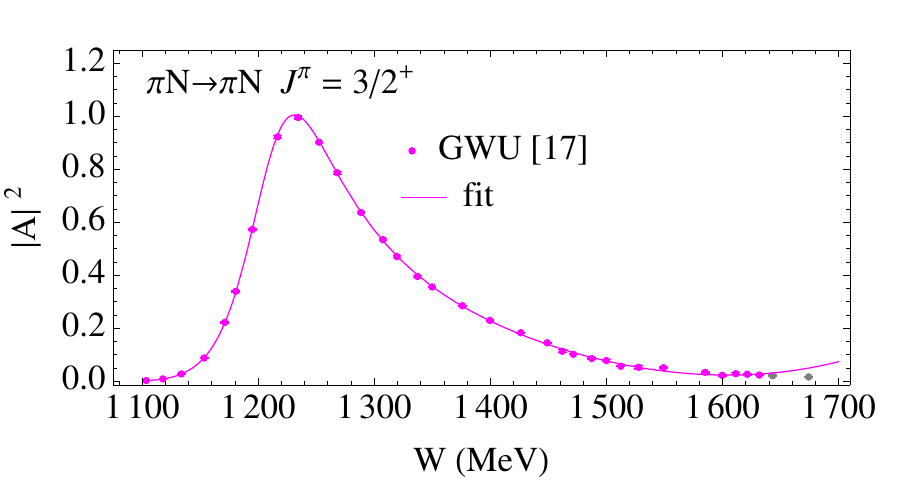}\\
\includegraphics[width=\figwidth]{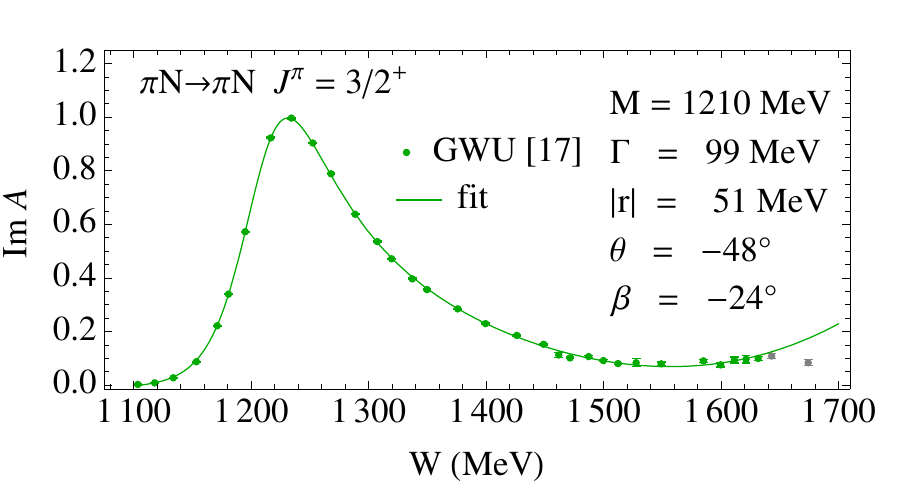}
\caption{Combined fit of $\mathrm{Im}\,A$ and $|A|^2$ obtained from Eq.~(\ref{AmplitudeFinal}) to the GWU $\pi N$ elastic $3/2^+$ partial wave from Ref.~\cite{Arn09}  \label{Delta1232piNelastic}}
\end{figure}

In Fig.~(\ref{Delta1232piNelastic}), we show a combined fit of $\mathrm{Im}\,A$ and $|A|^2$ to the GWU $\pi N$ elastic partial wave with $J^\pi =3/2^+$ from Ref.~\cite{Arn09}. 

\begin{figure} 
\includegraphics[width=\figwidth]{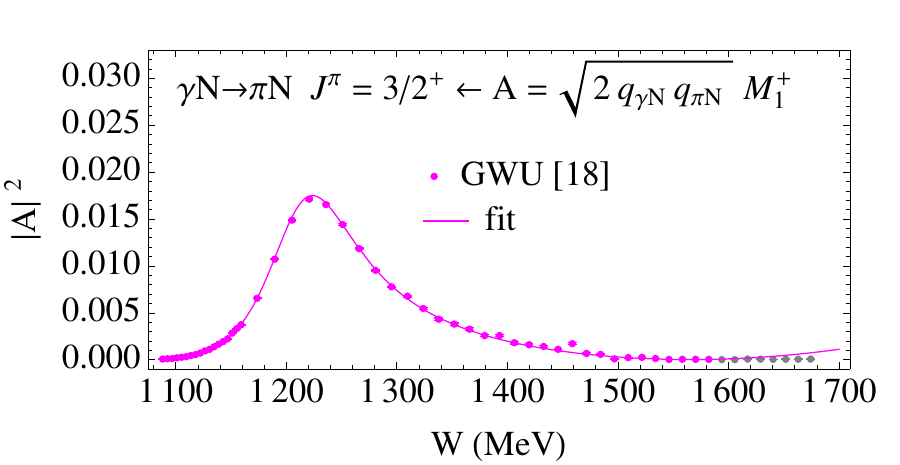}\\
\includegraphics[width=\figwidth]{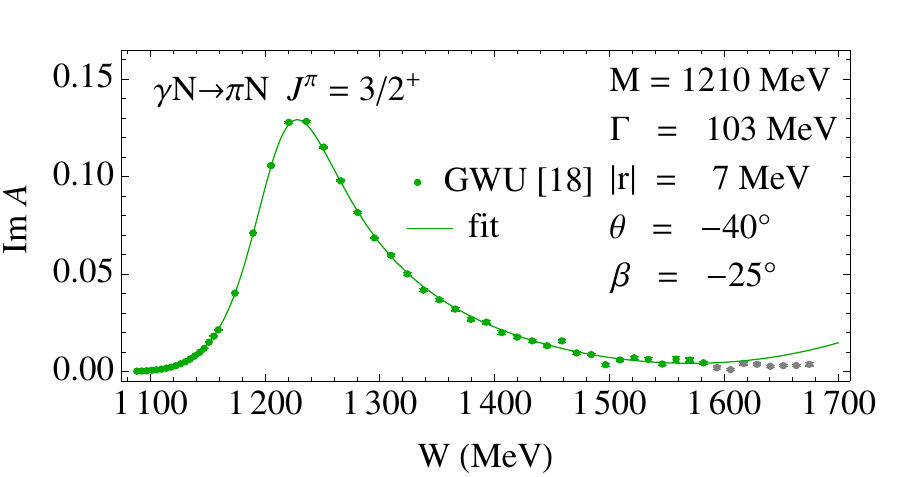}
\caption{Combined fit of $\mathrm{Im}\,A$ and $|A|^2$ obtained from Eq.~(\ref{AmplitudeFinal}) to the GWU $\gamma N\rightarrow \pi N$ $M_1^+$ partial wave taken from Ref.~\cite{SAID}\label{Delta1232gammaNpiNpM}}
\end{figure}

Finally, we do a combined fit of $\mathrm{Im}\,A$ and $|A|^2$ to the GWU $\gamma N\rightarrow \pi N$ partial wave $M_1^+$ from Ref.~\cite{SAID}, as shown in Fig.~(\ref{Delta1232gammaNpiNpM}). 

Extracted resonance parameters of $\Delta(1232)$ are shown in Table \ref{Delta1232Table}. Curiously, $\beta$ remains the same in various reactions, similarly to fundamental resonance parameters $M$ and $\Gamma$. Therefore, we investigate the properties of $\beta$.

\begin{table}[h!]
\caption{ Resonance parameters of $\Delta(1232)$ extracted from the data (generally, $|r|$ and $\theta$ vary for different processes). PDG estimates for $\pi N\rightarrow\pi N$ are shown for comparison.
\label{Delta1232Table}}
\begin{ruledtabular}
\begin{tabular}{lcccccc}
$\mathbf{\Delta(1232)}$ & $M$ & $\Gamma$ & $|r|$ & $\theta$ & $\beta$ & $M_\mathrm{BW}$  \\
Process	& (MeV) & (MeV) & (MeV) & ($^\circ$) & ($^\circ$) & (MeV) \\
 \hline
$\pi N\rightarrow \mathrm{all}$ & 1211$\pm$1 & 101$\pm$1 & 53$\pm$2 & $-46\pm$1 & $-25\pm$2& 1234$\pm$2  \\
$\pi N \rightarrow \pi N$ & 1210$\pm$1 & 99$\pm$1 & 51$\pm$1 & $-48$$\pm$1 & $-24\pm$1 & 1233$\pm$2  \\
$\gamma N \rightarrow \pi N$ & 1210$\pm$1 & 103$\pm$1 & 7$\pm$1 &  $-40\pm$1 & $-25\pm$1 & 1235$\pm$2  \\
\hline
PDG \cite{PDG} & 1210$\pm$1 & 100$\pm$2 & 52$\pm$1 & $-47$$\pm$1 & $-24$$\pm$1 & 1232$\pm$1 \\
\end{tabular}
\end{ruledtabular}
\end{table}

\section{Fundamental parameter \boldmath$\beta$}

The amplitude in Eq.~(\ref{AmplitudeFinal}) is, in fact, a matrix element of the full amplitude matrix $A_{ab}$ in channel indices $a$ and $b$. Each process from channel $a$ to channel $b$ (e.g.,~from $\gamma N$ to $\pi N$) is described by a corresponding matrix element 
\begin{equation}
A_{ab} = x_{ab} \, e^{i\left(\rho+\beta_{ab}\right)} \sin\left(\rho+\delta_{ab}\right).
\end{equation}  
When the unitarity constraint is imposed on the amplitude matrix, $A^\dag A=\mathrm{Im}\, A$, we immediately see that it cannot be satisfied unless all parameters $\beta_{ab}$ are the same for all processes. {\bf Consequently, the phase \boldmath$\beta$ must be the same in all processes due to unitarity.}

Since $\beta$ is invariant, we rewrite the amplitude as 
\begin{equation}\label{AmplitudeX}
\boxed{A = X \, e^{i(\rho+\beta)} \sin(\rho+\beta),}
\end{equation}
where $X_{ab}$ is a real symmetrical matrix given by $x_{ab}\,\sin\left(\rho + \delta_{ab}\right)/\sin(\rho+\beta)$. Imposing unitarity on this form drastically simplifies it and gives us $X^2=X$, i.e., that $X$ is a projection matrix. Since $X$ is symmetrical projection matrix, the trace of $X$ equals its rank. Therefore, we get
\begin{equation}
\mathrm{tr}A = \mathrm{rank}X\,\, e^{i(\rho+\beta)} \sin(\rho+\beta).
\end{equation}
Consequently, the trace of the amplitude's real part will be zero, while the trace of the imaginary part will have a maximum at the energy equal to the Breit-Wigner mass, since $\rho+\beta$ equals $90^\circ$ there. 

That is somewhat extraordinary because this result from a simple, non-analytic, single-resonance model with only a few free parameters is in perfect agreement with the result of a full-scale, analytic, unitary, coupled-channel, and mutiresonance analysis we did in Ref.~\cite{CecPLB08}. There, we have noted that these peaks and zeros curiously correspond to the Breit-Wigner masses of N$^*$ resonances, but also to the real poles of the K-matrix, which was all highly controversial at the time (see e.g.~Ref.~\cite{Arn09}). 

And what about the K-matrix within the model proposed here? The K-matrix is defined as \mbox{$A\, (I+i\,A)^{-1}$}, where $I$ is the unit matrix, from which we get
\begin{equation}
\boxed{K=X\,\tan(\rho+\beta).}
\end{equation} 
This $X$ is the same as the one in Eq.~(\ref{AmplitudeX}). Evidently, the K-matrix trace will have  poles at the Breit-Wigner masses, exactly as it was reported in Ref.~\cite{CecPLB08}. 



\section{Conclusions}

In conclusion, we propose the Breit-Wigner phase $\beta$ as the new fundamental resonance parameter because it is the same in all processes, directly connects the Breit-Wigner to the pole mass and width, shifts the complex phase of the resonant amplitude, and links the K-matrix poles with the Breit-Wigner masses. Therefore, we suggest that $\beta$ is added to resonance properties tables (perhaps instead of the Breit-Wigner mass and width).






\end{document}